\documentclass[twocolumn,pra,aps,showpacs]{revtex4}
\usepackage{xspace,amsmath,amsfonts,amsthm,amssymb,amsbsy,color,graphicx,epstopdf,soul}
\usepackage[breaklinks=true]{hyperref}
\hypersetup{
  colorlinks   = true, %Colours links instead of ugly boxes
  urlcolor     = blue, %Colour for external hyperlinks
  linkcolor    = blue, %Colour of internal links
  citecolor   = red %Colour of citations
}

\newcommand{\sq}[1]{\left[ {#1} \right]}
\newcommand{\tr}[1]{{\textrm {Tr}}\sq{#1}}
\newcommand{\smallfrac}[2]{\mbox{$\frac{#1}{#2}$}}
\newcommand{\half}{\smallfrac{1}{2}}
\newcommand{\bra}[1]{\langle{#1}|}
\newcommand{\ket}[1]{|{#1}\rangle}

\newcommand{\op}[2]{\ket{#1}\bra{#2}}
\newcommand{\enavg}[1]{\mathrm{E}\sq{#1}}
\newcommand{\gravg}[1]{\mathbb{E}\sq{#1}}
\newcommand{\expt}[1]{\langle{#1}\rangle}
\newcommand{\dg}{^\dagger}
\newcommand{\D}[1]{{\cal D}\sq{#1}}

\newcommand{\Hc}[1]{{\cal H}\sq{#1}}
\newcommand{\Hcun}[1]{\tilde{\cal H}\sq{#1}}

\newcommand{\nn}{\nonumber} 

\newcommand{\erf}[1]{Eq.~(\ref{#1})}
\newcommand{\frf}[1]{Fig.~\ref{#1}}
\newcommand{\srf}[1]{Sec.~\ref{#1}}

\definecolor{nblue}{rgb}{0.3,0.3,1.0}%229
\definecolor{ngreen}{rgb}{0.2,0.7,0.2}%161
\definecolor{nred}{rgb}{0.9,0.1,0}%711&900

\begin{document}
\title{Rapid readout of a register of qubits using open loop quantum control }%and continuous measurements}
\author{Joshua Combes$^{1,2,3,4}$}%\email{joshua.combes@gmail.com}
\author{Aaron Denney$^{1}$}
\author{Howard M. Wiseman$^3$}
\affiliation{$^1$Center for Quantum Information and Control, University of New Mexico, Albuquerque, NM 87131-0001, USA\\
$^2$Research School of Engineering, Australian National University, Canberra, ACT 0200, Australia\\
$^3$Centre for Quantum Computation and Communication Technology (Australian Research Council), Centre for Quantum Dynamics, Griffith University, Brisbane, Queensland 4111, Australia\\
$^4$Centre for Engineered Quantum Systems, School of Mathematics and Physics, The University of Queensland, St Lucia, QLD 4072, Australia}

\begin{abstract}
Measurements are a primitive for characterizing quantum systems. Reducing the time taken to perform a measurement may be beneficial in many areas of quantum information processing. We show that permuting the eigenvalues of the state matrix in the logical basis, using open loop control, provides a $O(n)$ reduction in the measurement time, where $n$ is the number of qubits in the register. This reduction is of the same order as the (previously introduced) locally optimal feedback protocol. The advantage of the open loop protocol is that it is far less difficult experimentally. Because the control commutes with the measured observable at all times, our rapid measurement protocol could be used for characterising a quantum system, by state or process tomography, or to implement measurement-based quantum error correction.

 % we give a explict fixed quantum circuit that implements the open loop control with $O(log n)$ cnots. 
\end{abstract}

\pacs{03.65.Yz,3.65.Wj,42.50.Lc,02.30.Yy}

\maketitle

%\tableofcontents

Quantum measurements are typically treated as instantaneous but in many practical situations this is clearly not the case. The noisy signals produced by the detectors need to be integrated over a time interval, called the measurement time, in order to determine the measurement outcome. For example spin qubits made from GaAs double dot systems have measurement times of the order of $\sim\!\!10\mu$s~\cite{gasdqd}, which is much longer than the $n$s timescale of the internal dynamics. 

Schemes for characterizing quantum systems require a large number of such measurements. The process of characterization can be lengthy for a number of reasons. In quantum state tomography it is necessary to repeatedly produce a state, choose a measurement basis, perform the measurement, and then process all the results into a state estimate. 
Each of these process can be time consuming. One approach to reduce the time required to characterize a system, provided it is low rank, is to perform compressed sensing \cite{GroFlamBec10} or direct fidelity estimataion \cite{FlaLui11,daSlLanPou11} instead of tomography. Compressed sensing for example, under some reasonable assumptions, drastically reduces: the number of measurements, the post-processing of the data and hence the total characterization time even if the measurements are slow. 

A reduced characterization time can also be achieved by speeding up the measurement process. It has been found previously that by applying control throughout the measurement process it is possible to affect the measurement rate \cite{ComWisJac08}. There are two important ingredients in searching for procedures which speed up the measurement process. The first is a description of  the measurement process that includes time. This ingredient is the quantum trajectory description of the measurement process. The second ingredient is restricting the allowed controls to a set of operations that 
 make the measurement projective, and allow us to retrodict the basis of the measurement.  The relevant restriction on the control is to operations that commute with the measured observable. In other words, we wish to restrict the control to permutations in the initial measurement (logical) basis~\cite{ComWisJac08}.

\begin{figure}[h]
\begin{center}
\leavevmode \includegraphics[width=0.65\hsize]{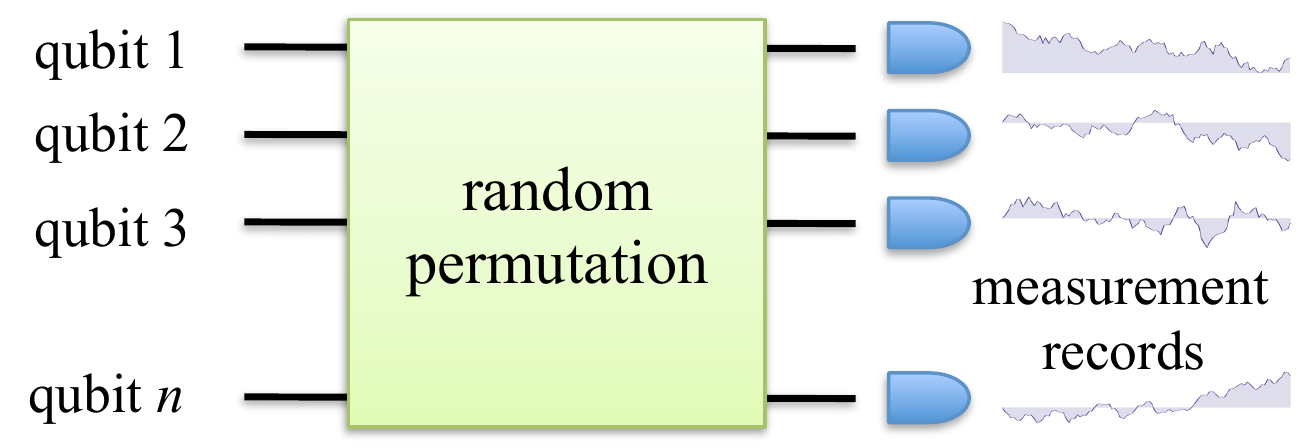}
\caption{In our protocol $n$ qubits are independently and continuously monitored in the logical basis. Open loop control consisting of random permutations in the logical basis are applied throughout the measurement to speed up the wavefunction collapse. As the control commutes with the measurements our protocol is comparable with state tomography.}\label{fig1} 
\end{center}
\end{figure} 

Reference~\cite{ComWisJac08} introduced a feedback protocol, optimised locally in time, that reduces the measurement time by a factor that scales linearly in the number of qubits, $n$, in a register. The ratio of the  measurement time without control to that with control is called the speed-up \cite{Jac0303}. It was found that the speed-up was $S=0.7n$. Unfortunately quantum feedback is experimentally demanding. Furthermore the protocol suggested in Ref.~\cite{ComWisJac08} requires a rapidly reconfigurable quantum circuit with a large number of CNOT gates. Both of these factors suggest the protocol in Ref.~\cite{ComWisJac08} may not be practical to implement.

In this article we present open loop control protocol, inspired by Ref.~\cite{ComWisSco10}, which achieves the same $O(n)$ improvement in speed-up. The protocol consist of randomly permuting the state in the measurement eigenbasis throughout the measurement; see \frf{fig1}. The expected asymptotic speed-up of this protocol is $S\sim 0.5n$ and simulations for small $n$ show a $S\sim 0.4n$ scaling. We emphasize that the continuous time description is not necessary to apply our results in practice. Simply permuting the state in the logical basis throughout the measurement and retrodicting the final result is sufficient. This protocol should be contrasted to rapid purification protocols which can not be used in a tomographic setting \cite{Jac0303,ComWisJacOoc10,ComWis11}

This article is organized as follows. In \srf{sec_cm}, we determine the rate at which information is extracted from a register of qubits using the quantum trajectory description of a continuous measurement. Next, we review the locally optimal protocol feedback protocol, with controls restricted to the permutation group, for rapid measurement in \srf{sec_lo}. We then reconsider the rapid-measurement problem with the open loop controls restricted to the permutation group in \srf{sec_rp}. We show analytically that the scaling of the speed-up with the number of qubits register is upper and lower bounded by a quantity that scales with $n$. Numerical simulations for small $n$ in \srf{nan} show the scaling is $S\sim 0.4n$. We conclude in \srf{dis} with a discussion of open problems.

%===============================================
\section{continuous measurement}\label{sec_cm}
%===============================================
In this section we derive the rate at which a continuous measurement extracts information from a quantum system for a reasonably general measurement model. The starting point for our analysis is the quantum trajectory description of the measurement process. 

The physics behind the quantum trajectory description is as follows. Consider using an ancilla, e.g. a field or a current, to probe a quantum system. The interaction between the system and the probe correlates the two systems. Typically the interaction only weakly correlates the system and ancilla. By measuring the ancilla and then using quantum measurement theory we can determine the state of the system conditioned on the results of the measurement. In order to obtain a strong measurement many ancilla must be coupled and then measured. In the limit where there is a continuum of ancilla the measurement process becomes continuous in time. The quantum trajectory approach is a way to describe the conditional state of the system as a functional of the measurement record $R(t)$. One uses a stochastic differential equation for the state of the system $d\rho_{R(t)}$  to describe the change of the system state over an infinitesimal time interval due to any unitary, dissipative and measurement induced dynamics. The system state at the next time interval is given by  $\rho(t+dt)=\rho(t) +d\rho_{R(t)}$. Interestingly the form of the differential equation for the system state is similar across many different systems \cite{GamBlaBoi08,JorKor06,Oxtoby}. Accessible derivations of this process can be found in Refs.~\cite{Brun02,JacSte06,WisMil10,filterref}.

\subsection{continuous measurement of a single qubit}
Consider a finite-dimensional quantum system undergoing a continuous measurement of an observable $X$.  The change to our state of knowledge of an individual system, $\rho$, conditioned on the result of the measurement in an infinitesimal interval is described by the stochastic master equation (SME) ~\cite{Brun02,JacSte06,WisMil10} 
\begin{eqnarray}\label{SME}
 d\rho[t;X]= 2\gamma \, dt\, \D{X}\rho(t) +\sqrt{2\gamma}\,dW(t)\,\Hc{X}\rho(t),  
  \end{eqnarray}
where $\D{A} \rho \equiv A\rho A\dg -\half (A\dg A \rho + \rho A\dg A)$ and $\Hc{A} \rho \equiv  A\rho +\rho A\dg - \tr{(A\dg+ A )\rho}\rho$. Here we are working in a frame that removes any Hamiltonian evolution. The {\em measurement strength}, $\gamma$, determines the rate at which information is extracted. The measurement  result in the interval $[t,t+dt)$ is  
\begin{eqnarray}\label{dR}
 dR = 2\sqrt{2\gamma}\langle X(t) \rangle dt + dW(t), 
\end{eqnarray}
 where $dW$ is a Wiener process and $\expt{X(t)} = \tr{X\rho(t)}$. Without loss of generality we take $X$ to be traceless. We can do this because \erf{SME} is invariant under $X \to X+\lambda I$ for $\lambda \in \mathbb{R}$ where $I$ is the identity operator.

\subsection{continuous measurement of a register of qubits}
We now generalize the SME in \erf{SME} to a register of $n$ qubits, where each qubit is independently and continuously measured. Instead of one observable $X$, we now have $n$, given by 
\begin{align}
Z^r= I^{(1)}\otimes I^{(2)}\otimes \ldots \sigma_z^{(r)}\ldots \otimes I^{(n)},
\end{align} 
where $r$ labels the $r$th qubit. The SME describing such a measurement is
\begin{equation}\label{SME_reg}
  d\rho = \sum _{r=1}^{n} 2{\gamma} \, dt\,\D{Z^{r}}\rho+\sqrt{2{\gamma}}\,dW^{(r)}\Hc{Z^{r}}\rho,
\end{equation}
and the $n$ Wiener processes obey $dW^{(i)}dW^{(j)}=\delta _{ij}dt$.

We would like to quantify the amount of information the measurement provides as a function of time. Thus we must choose a measure of information. %The von-Neuman entropy seems like the obvious choice however it is a nonlinear function of the state. This becomes a problem when trying get closed form expressions for the time evolution. 
Previous research has used the logarithm of the infidelity, the log-infidelity \cite{ComWisJac08}, which is defined as $\ln\Delta$, where $\Delta=1-\lambda_{0}$, where $\lambda_0$ is the largest eigenvalue of $\rho$. This measure has two advantages: (1) it is often possible to get approximate closed form solutions for the log-infidelity and (2) the time $\tau$ at which the average log-infidelity $\expt{\ln(\Delta(\tau))}$ reaches a fixed log-infidelity $\ln{\epsilon}$ is related to the mean time $\expt{T}$ to reach that infidelity $\epsilon$ \cite{WisRal06,ComWisJac08}.

In order to proceed in our analysis of information extraction rates in a register of qubits we need to formulate a solution to \erf{SME_reg}. This requires the machinery of linear quantum trajectories \cite{JacSte06,WisMil10}. The linear and consequently unnormalised version of \erf{SME_reg} is
\begin{equation}\label{SME_reglin}
  d\tilde{\rho} = \sum _{r=1}^{n} 2{\gamma} \, dt\,\D{Z^r}\tilde{\rho}+\sqrt{2{\gamma}}\,dR^{(r)}\Hcun{Z^r}\tilde{\rho},
\end{equation}
where $\Hcun{A}B= AB+B A\dg$ and the tilde denotes the lack of normalisation, 
and $dR^{(r)}$ is related to $dW^{(r)}$ analogously to $dR$ in \erf{dR}.  
 The linear trajectory solution for this equation is
\begin{align}
\nn \tilde{\rho}({\bf R},t)=& e^{-2{\gamma} I t}e^{\sqrt{2{ \gamma}}Z^n R^n}\ldots e^{-2{ \gamma} I t}e^{\sqrt{2{ \gamma}}Z^1 R^1}\\
&\rho(0) e^{-2{ \gamma} I t}e^{\sqrt{2{ \gamma}}Z^1 R^1}\ldots e^{-2{ \gamma} I t}e^{\sqrt{2{ \gamma}}Z^n R^n }\!,
\end{align}
where $R^{ r}=\int_{0}^{t}dR^{ (r)}$, and $\bf R$ is the vector of records $(R^1, R^2, \ldots, R^n)$. We have also taken the initial state to be $\rho(0)= I/2^{n}$, corresponding to no information about the system. This assumption 
is not necessary for the asymptotic results below, but makes the derivation more elegant. The linear trajectories expression for the normalization factor of this state is
\begin{eqnarray}
  \mathcal{N}({\bf R},t)=\tr{\tilde{\rho}({\bf R},t)}
    &=& \frac{e^{-4{\gamma} n t}}{2^n}\sum_{q=0}^{2^n-1}e^{2\sqrt{2{\gamma}} \mathcal{R}^q}.
\end{eqnarray}
Here the $\mathcal{R}^q$ are linear combinations of the $n$ ``bare'' records $R^{r}$. The records $\mathcal{R}^q$ hide a lot of the complexity of this expression. Thankfully the linear combinations are simply all the $2^{n}$ possible combinations of adding and subtracting the $n$ bare records. The index $q$ is actually an $n$ digit binary string which describes how the records $\mathcal{R}^q$ are constructed from the bare records. A `0' in the $q$th position corresponds to a plus coefficient in front of the $r$th bare record, for $r=q+1$ (recall that $q\in \{0,\ldots, 2^{n}-1\}$). A `1' in the $q$th position corresponds to a minus coefficient in front of that bare record. For example in a two qbit register the four $\mathcal{R}^q$'s would be $\mathcal{R}^{00}=R^{1}+R^{2}$, $\mathcal{R}^{01}=R^{1}-R^{2}$, $\mathcal{R}^{10}=-R^{1}+R^{2}$, and $\mathcal{R}^{11}=-R^{1}-R^{2}$. 

The normalized expression for the conditional state is $\rho({\bf R},t) = \tilde{\rho}({\bf R},t)/ \mathcal{N} $, which will always be diagonal in the logical basis. Thus, without loss of generality, 
we can define the local qubit bases (and the sign of the measurement records) such that $\lambda_{\bar{0}}$, the largest eigenvalue of $\rho$, corresponds to $\ket{\bar{0}}\equiv\ket{00\ldots0}$. The expression for the largest eigenvalue is
\begin{equation}
\lambda_{\bar{0}}= \frac{ \exp{\left ({2\sqrt{2{ \gamma}} (R^1+R^2\ldots +R^r\ldots +R^n)}\right )}  } 
                           { \sum_{q=0}^{2^n-1}\exp{(2\sqrt{2{ \gamma}} \mathcal{R}^q)}  },
\end{equation}
here $\lambda_{\bar{0}}$ does not have the records ${\bf R}$ as an argument for notational compactness. 
It is known that measurement tends to cluster the large eigenvalues around the largest eigenvalue \cite{ComWisJac08} (also see the appendix of \cite{ComWis11}). That is, eigenvalues that are close to the largest eigenvalue contain most of the probability. This is graphically depicted in \frf{fig2} b. Thus we only consider eigenvalues one Hamming unit away from the largest eigenvalue $\lambda_{\bar{0}}$ and truncate the normalization to a total of $(n+1)$ terms. With this approximation and in the long-time limit the largest eigenvalue is 
\begin{equation}
\lambda_{\bar{0}} =  \frac{e^{2\sqrt{2{\gamma}} (R^1+\ldots R^n)}}{e^{2\sqrt{2{\gamma}} (R^1+\ldots  R^n)}+\sum_{r=1}^{n}e^{2\sqrt{2{\gamma}}((R^1+\ldots R^n)-2 R^r)}}. 
\end{equation}
which simplifies to %\st{Dividing the numerator and denominator by the numerator and simplifying gives}
\begin{equation}
\lambda_{\bar{0}}  =    \frac{1}{1+\sum_{r=1}^{n}e^{-4\sqrt{2{\gamma}}R^r} }.
\end{equation}
When $t\gg {\gamma} ^{-1}$ the bare records will, on average, satisfy $R^1\approx R^2\approx \ldots R^n\approx 2\sqrt{2{\gamma}} t$  (because $dR^r \approx 2\sqrt{2{\gamma}} dt +dW$) so that 
\begin{equation}
\lambda_{\bar{0}} \approx  \frac{1}{1+ne^{-4\sqrt{2{ \gamma}}(2\sqrt{2{ \gamma}} t)} }.
\end{equation}
Now that the expression for the largest eigenvalue is suitably simple, we can work out $\ln{(1-\lambda_{\bar{0}} )}$:
\begin{equation}
\ln{(1-\lambda_{\bar{0}} )} %=\ln{\frac{ne^{-4\sqrt{2\gamma}c}}{1+ne^{-4\sqrt{2\gamma}c} }}
\approx \ln{ \big( ne^{-16{ \gamma} t}\big) }=-16{ \gamma} t+\ln{n}.
\end{equation}
The long-time expression for $\expt{\ln\Delta}$ in the absence of feedback is thus 
\begin{align} 
 \expt{\ln[\Delta(t)]}_{\rm nfb}\sim -16{\gamma} t. \label{nfbLongTimeLimit}
\end{align}   
From this relation we expect that the mean time to attain infidelity $\Delta=\epsilon$ is, for $\ln(\epsilon^{-1}) \gg 1$, \cite{WisRal06,ComWisJac08}
\begin{eqnarray} \label{mtreg}
\expt{T}_{\rm nfb} = (1/16{\gamma})\ln(\epsilon^{-1}) .
\end{eqnarray}
An astute reader will notice that the factor $\ln n$ did not turn up in \erf{mtreg}, this is because it is negligible at long times as it scales as  $\ln n /t$.

\begin{figure}[h]
\begin{center}
\leavevmode \includegraphics[width=0.85\hsize]{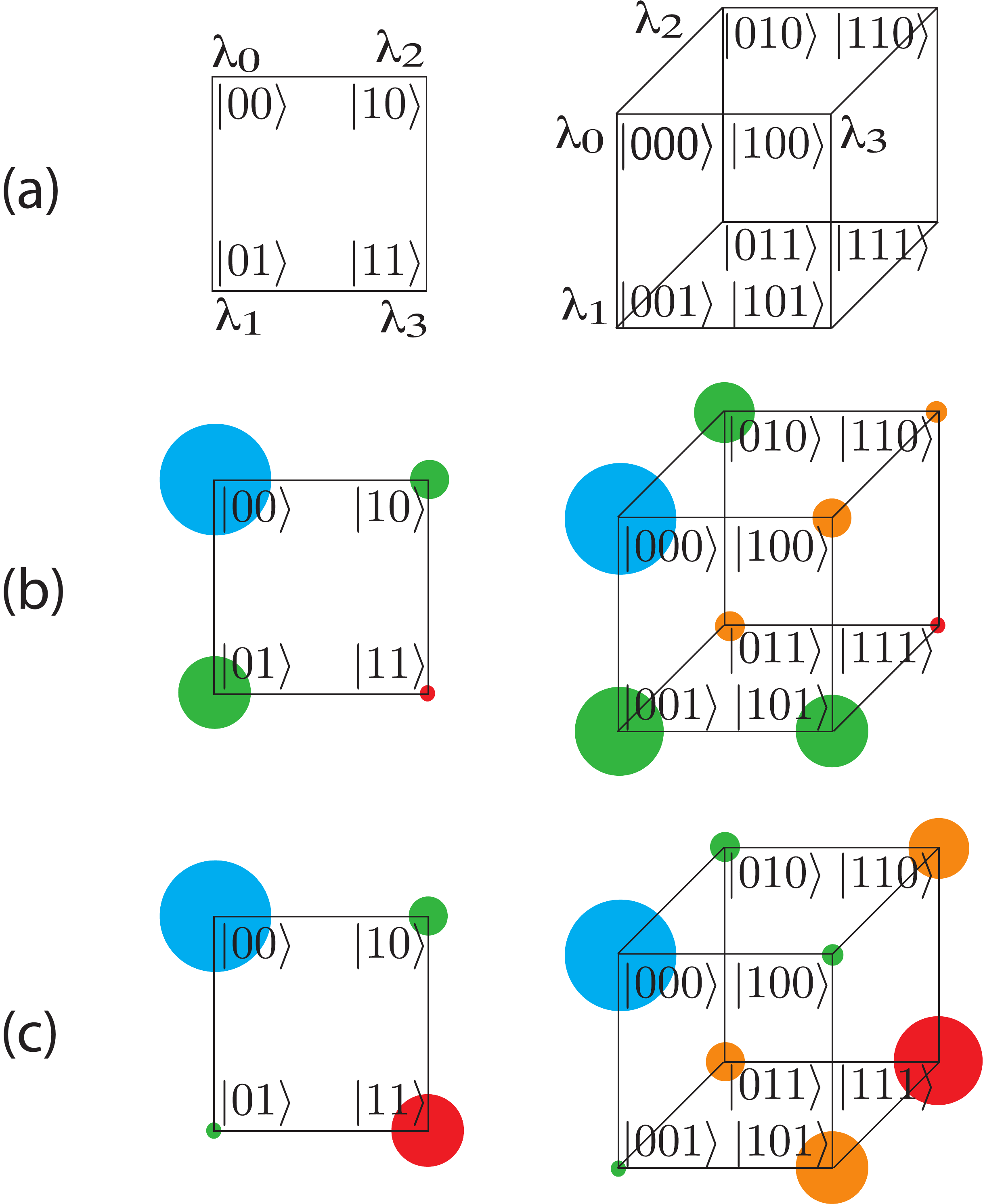}%.eps
\caption{To gain some intuition about the measurement process in a register we illustrate the effect of measurement on the eigenvalues of $\rho$ by plotting them on a Hamming cube. A Hamming cube is a way of depicting the distance between $n$ bit binary strings. The distance between two strings is given by the number of bit flips required to get from one string to the other. To plot the state matrix on the Hamming cube we associate the eigenstates with the vertices of the cube and the eigenvalues are placed on the near the vertices as in (a). We used the convention that the largest eigenvalue is always labeled $\lambda_0$.  By appropriately relabelling the eigenstates one can always think of the largest eigenvalue being associated with the eigenstate $\ket{\bar{0}}$.
In figure 1 (b) we plot how a continuous measurement affects the eigenvalues. We now represent the eigenvalues as circles. The magnitude of the eigenvalues is denoted by the size of the circle, while the distance in Hamming space is given by colour. For example blue and red are maximally distant in color space and so are the corresponding eigenvalues. From figure (b) we conclude that measurement tends to clump the large eigenvalues close to the largest eigenvalues. In figure (c) we depict the H-ordered state which is locally optimal for rapid measurement. Our open loop protocol randomly permutes the eigenvalues over all vertices of the Hamming cube.}\label{fig2} 
\end{center}
\end{figure}

%===============================================
\section{locally optimal rapid measurement}\label{sec_lo}
%===============================================
In this section we briefly review the rapid measurement protocol developed in Ref.~\cite{ComWisJac08}. The basic procedure is to derive a stochastic differential equation for the log-infidelty and then choose a control that maximizes the reduction in the log-infidelty.

Starting with the SME \erf{SME_reg} we transform the observable $Z^r$ to $ Z^r-I$. We order the eigenvalues of $\rho$ such that $\lambda_\alpha\ge\lambda_\beta$ when $\alpha<\beta$, where $\alpha, \beta$ are binary strings, i.e. $\lambda_{00\ldots0}\ge\lambda_{00\ldots01}\ge\ldots\ge\lambda_{11\ldots1}$. Again we assume the largest eigenvalue of $\rho$, i.e. $\lambda_{\bar{0}}$, is placed such that it corresponds to $\ket{\bar{0}}=\ket{00\ldots0}$. Then 
from \erf{SME_reg} the equation of motion for an arbitrary population of $\rho$, $i$, in the logical basis is 
 \begin{equation}
  d\lambda_i =2\sqrt{2\gamma}\sum _{r=1}^{n} dW^{(r)}(\,\expt{i|Z^r|i}-\tr{Z^r\rho}\,)\lambda_i, \label{dlambda}
\end{equation}
where we have used $\D{Z^r}\rho = 0 \,\, \forall r$. The equation for the largest eigenvalue $d\lambda_{ \bar{0} }$ is
\begin{eqnarray}
  d\lambda_{\bar{0} } &=& -2\sqrt{2\gamma}\lambda_{\bar{0}}\sum _{r=1}^{n} dW^{(r)}\tr{Z^r\rho}.
\end{eqnarray}
Using the stochastic change of variables rule (i.e. Ito's lemma) the SDE for the log-infidelty is
\begin{align}
\nn d\ln{(1-\lambda_{\bar{0} })}&=d\lambda _{\bar{0} } d\ln{(1-\lambda_{\bar{0} })} + \half (d\lambda _{\bar{0} })^2 d^2\ln{(1-\lambda_{\bar{0} })} \\
&=\frac{-(d\lambda _{\bar{0} })^2}{2(1-\lambda_{\bar{0} })^2}+\frac{-d\lambda _{\bar{0} }}{(1-\lambda_{\bar{0} })} 
\end{align}
Thus to complete the change of variables we also need to work out the $(d\lambda _{\bar{0}})^2$ term. Because it is more complicated we write it in full:
\begin{eqnarray}
  (d\lambda_{\bar{0}})^2 &=& (-2\sqrt{2\gamma}\lambda_{\bar{0}})^2\sum _{r,s} dW^{(r)}dW^{(s)}\expt{Z^r}\expt{Z^s}\nn \\
&=&8\gamma\lambda_{\bar{0}}^2dt\sum _{r}\expt{Z^r}^2,
\end{eqnarray}
where we have used the fact that $dW^{(i)}dW^{(j)} =\delta _{ij}dt$. Finally we find the average rate of change of the log-infidelity to be \cite{ComWisJac08}
\begin{eqnarray}
\enavg{d \ln{\Delta} }&=&-4{\gamma} \,dt\sum _{r}\expt{Z^r}^2\frac{(1-\Delta)^2} {\Delta^2}.
\end{eqnarray}

Now we wish to maximize the average reduction of the log-infidelity for a given $\rho$ using feedback. This is achieved by reordering the elements of $\rho$, so as to maximize $\sum_r\expt{Z^r}^2$, in the following way.  By definition (above) the largest eigenvalue is at $\ket{\bar{0}}$. The second largest eigenvalue $\lambda_{00\ldots01}$ is then placed at $\ket{\bar{1}}$ such that it is the maximum Hamming distance \cite{Ham1950} away. The next $n$ largest eigenvalues are placed at one Hamming unit away from  $\ket{\bar{1}}$, the next $^nC_2$ largest eigenvalues are placed two Hamming units away from $\ket{\bar{1}}$, and so on. This ordering has been called H-ordering~\cite{ComWisJac08}. Example H-orderings for a two- and three-qubit register are depicted in the  Fig.~\ref{fig2} c. A feedback protocol which H-orders at every time instant is said to be locally optimal (LO) in time. This should be contrasted to globally optimal protocols which require optimal control tools like dynamic programming~\cite{WisBou08,TeoComWis12}. 

We now bound, for a register of qubits, the amount by which the H-ordering algorithm speeds up the measurement process. To do this we must bound  
\begin{equation}
\sum_r\expt{Z^r}^2.\label{zsum}
\end{equation}
Consider an arbitrary state $\rho$ with an infidelity $\Delta$. The upper bound on \erf{zsum} is obtained by considering the minimally mixed state with the same infidelity: $\rho_{2}=\mathrm{diag}(1-\Delta,\Delta,0,\cdots,0)$ \cite{ComWisJac08}. H-ordering this state corresponds to $\lambda_{0\ldots 0} = 1-\Delta$ placed at $\ket{\bar{0}}$ and the other eigenvalue $\lambda_{0\ldots 01} = \Delta$ is placed at $\ket{\bar{1}}$. The lower bound on \erf{zsum} is obtained by considering the state: $\rho_{F}=(1-\Delta, \delta, \cdots, \delta)$ where $\delta = \Delta/(2^{n}-1)$.  H-ordering this state corresponds to placing $\lambda_{0\ldots 0} = 1-\Delta$ at $\ket{\bar{0}}$ and distributing the residual probability equally over the remaining $2^n-1$ eigenvalues. The reason why $\rho_{2}$ and $\rho_{F}$ provide upper and lower bounds on \erf{zsum} can be understood as follows. The feedback (H-ordering) is a permutation of the basis of $\rho$. Recall that the largest eigenvalue of $\rho$ is fixed at $\ket{\bar{0}}$. The state $\rho_{2}$ is very sensitive to permutations, with respect to \erf{zsum}. Conversely the state $\rho_{F}$ is invariant under permutations of the remaining $2^{n}-1$ eigenvalues [with respect to \erf{zsum}]. 

The using the states $\rho_{2}$ and $\rho_{F}$ the it is simple to calculate the bounds on \erf{zsum}. The bounds are 
\begin{align}\label{boundsDelta}
[n2^{2n}/(2^n-1)^2] \Delta^2 \leq\sum_r\expt{Z^r}^2\leq 4n\Delta^2.
\end{align}
 In the long-time limit (i.e when $t\gg {\gamma} ^{-1}$ or equivalently $\Delta \ll1$) the bounds on log-infidelity are 
\begin{align}
\enavg{d\ln\Delta}_{\rm LO}=- 16{\gamma} dt  S_{\rm LO},\label{CLOlnDreg}
\end{align}
 where $S_{\rm LO}$ is bounded [from \erf{boundsDelta}] as
\begin{align}
 \frac{2^{2n}}{(2^n-1)^2} \frac{n}{4}  \leq S_{\rm LO}\leq  n . \label{SLUBreg} 
\end{align}
The solution to \erf{CLOlnDreg} is bounded as
\begin{align}
\enavg{\ln\Delta(t)}_{\rm LO}=-{16}{\gamma} t  S_{\rm LO},\label{CLOlnDregSOL}
\end{align}
In the long-time limit we cacluate a speedup factor from Eqs. (\ref{nfbLongTimeLimit}) and (\ref{CLOlnDregSOL}). Denoting the time taken by a measurement without feedback to reach a given value of $\ln\Delta$ as $t_{\textrm{nfb}}$, and that for the feedback protocol as $t_{\textrm{fb}}$, we equate $\ln\Delta(t_{\rm fb})=\ln\Delta(t_{\rm nfb})$ and solve for the ratio $t_{\rm fb}/t_{\rm nfb}$. In the long time limit we define the speed up to be $S=(t_{\rm fb}/t_{\rm nfb})^{-1}$. Doing so gives 
$S = S_{\rm LO}$. That is, we have upper- and lower-bounded the asymptotic speedup for the Locally Optimal 
closed-loop control strategy. 
% is precicely the asymptotic speedup.}

For $n\gtrsim 7$ the lower bound on $S_{\rm LO}$ is well approximated by $n/4$. Technically are bounds on the speed-up factor with respect to the $\langle \ln\Delta \rangle$ measure. It has been shown that these  bounds well-approximate the behavior of the mean time $\langle T \rangle$ to a fixed infidelity and numerical results confirm the $O(n)$ scaling predicted in \erf{SLUBreg} ~\cite{ComWisJac08}. The scaling found in the numerics of Ref.~\cite{ComWisJac08} was
\begin{equation}\label{S_RM_reg_fit}
S_{\mbox{\scriptsize H}} = 0.718 n.
\end{equation}

%============================================================================
\section{Rapid measurement using Open loop control with Random Permutations}\label{sec_rp}
%============================================================================
Inspired by Ref.~\cite{ComWisSco10} we will now replace the locally optimal feedback control from \srf{sec_lo} with open loop control and quantum filtering. That is we combine the conditional evolution in \erf{SME_reg} with a control strategy that does not depend on the conditional state. Since the LO feedback is simply a permutation of $\rho$'s basis we will consider the effect of randomly permuting this basis at each instant in time. Working in the Heisenberg picture with respect to the control unitary, this is equivalent to permuting the basis of the observable.

Randomly permuting $Z_{r}$'s basis at each instant the SME becomes
\begin{equation}\label{SME_reg_rand}
  d\rho = \sum _{r=1}^{n} 2{\gamma} \, dt\,\D{P Z^rP }\rho+\sqrt{2{\gamma}}\,dW^{(r)}\Hc{PZ^rP}\rho
\end{equation}
where the permutation $P$ is independently drawn from the permutation group ${\mathfrak P}(2^{n})$ at each time interval. Since $P$ is chosen uniformly from  ${\mathfrak P}(2^{n})$ for this protocol we explicitly average over all permutations in a single time step to find an analytical approximation to the average speed-up. We denote the average over permutations symbolically by $\gravg{.}$. Averaging \erf{SME_reg_rand} over ${\mathfrak P}(2^{n})$ we find 
\begin{align}\label{SME_reg_rand_avg}
\gravg{d\rho} = \sum^{2^n}_{ s=1}\sum _{r=1}^{n}&\, \Bigg (
   \frac{2{\gamma}}{D!} \, dt\,\D{P_{s} Z^rP_{s} }\rho \,+ \nn \\
   &\sqrt{\frac{2\gamma}{D!}}\,dW^{(r)}_{\,(s)}\Hc{P_{s}Z^r P_{s}}\rho \Bigg  )
\end{align}
where $\gravg{\rho(t+dt)}= \gravg{\rho(t)} +\gravg{d\rho}$. The equation  of motion for an arbitrary population $\mathbb{P}_{i}$ is given by $d \mathbb{P}_{i}= \tr{\gravg{d\rho} \Pi_{i}}$ where $\Pi_{i}=\op{i}{i}$ and $\mathbb{P}_i(0) = \tr{\rho(0)\Pi_i}$. The equation analogous to \erf{dlambda} is
 \begin{equation}
  d  \mathbb{P}_{i}=2\sqrt{\frac{2\gamma}{D!}}\sum_{s=1}^{D!}\sum _{r=1}^{n} dW^{(r)}_{\,(s)}\bigg(\expt{i|Z^{r}_{s}|i}-\tr{Z^{r}_{s}\gravg{\rho}}\bigg) \mathbb{P}_{i}, \label{dlambda_rand_perm}
\end{equation}
where $dW_{j}^{\mu}dW_{i}^{\nu}= \delta_{j,i,\mu,\nu}dt$. The average log-infidelity becomes
 \begin{equation}
  \enavg{d\ln(1- \mathbb{P}_{0})} =-(d\mathbb{P}_{0})^{2}/ [2(1-\mathbb{P}_{0})^{2}].
\end{equation}
Recall that  \erf{SME_reg_rand} is invariant under the transformation $Z^{r} \rightarrow Z^{r}-I$. Making that transformation we find that $\expt{0|Z^{r}_{s}|0}=0$, which gives
\begin{equation}
  \enavg{d\ln(1- \mathbb{P}_{0})}=-\frac{4\gamma \,dt}{D!}\frac{(1-\Delta)^2} {\Delta^2}\sum_{s}\sum _{r}\expt{Z^r_{s}}^2.\label{loginfide}
\end{equation}
Equation \ref{loginfide} is exact, and now we try and derive bounds on this expression. As before we only need to place bounds on $\sum_{s}\sum _{r}\expt{Z^r_{s}}^2$ so we substitute the state $\rho_{2}$ and $\rho_{F}$ into \erf{loginfide}, keeping terms $O(\Delta^{2})$, to obtain the upper and lower bounds. Further steps are outlined in appendix \ref{appbound}. In the long-time limit (i.e when $t\gg \gamma ^{-1}$ or equivalently $\Delta \ll 1$) the bounds on log-infidelity for random permutation (RP) control are
\begin{align}  
\enavg{d\ln\Delta}_{\rm RP}=-16\gamma dt  S_{\rm RP},\label{lnDRPlnreg}
\end{align}
where the bounds on $S_{\rm RP}$  
\begin{align} 
\frac{2^{2n}}{\left(2^n-1\right)^2} \frac{n}{4}  \leq S_{\rm RP}\leq  \frac{2^{n-1}}{2^{n}-1}n.  \label{RPLUBreg} 
\end{align} 
Following the argument in the preceding section, we can identify $S_{\rm RP}$ with the speed-up, 
in terms of the average time for $\Delta$ to attain a given value $\epsilon \ll 1$, relative to the 
no-control case. Notice that the lower bound coincides with the lower bound given in \erf{SLUBreg}. For $n\gg 1$ the bounds become
\begin{align} 
0.25 n\leq S_{\rm RP}\leq  0.5n. \label{RPLUBregsimp} 
\end{align}

%============================================================================
\section{numerical analysis}\label{nan}
%============================================================================
While \erf{RPLUBregsimp} tells us that the speed up is $\Theta(n)$ \footnote{The notation $X = \Theta(Y )$ means $X$ is both upper and lower bounded by quantities scaling as $Y$.} it does not tell us if the speed up is closer to the upper or lower limit. For this reason we perform numerical simulations and calculate the speed up and compare it to the bound. 

Our simulation method is to solve equation \erf{SME_reg} while applying a single random permutation at every time step. Then at each time step we calculate $(1-\lambda_{\rm max})$. By running many simulations we can build up statistics of the mean time to reach a fixed infidelity. The ratio of the mean time without control to the time with control is the speed up.
 
 In \frf{Fig6} we plot the numerically calculated speed-up for two- and three-qubit registers that are subject to continuous monitoring and random permutation open loop control. In both cases the calculated speed-up is greater than one. The dashed red lines are our numerically estimated speed ups in the limit that $\Delta\rightarrow 0$. These lines are calculated as follows. We take the mean time to infidelities between $10^{-4}$ and $10^{-6}$, for the no control and control cases, and perform linear regression to determine the slope. The ratio of these slopes is the expected speed up in the mean time to an infidelity of zero, which we call the {\em asymptotic speed up}.
  
\begin{figure}[h]
\begin{center}
\leavevmode \includegraphics[width=\hsize]{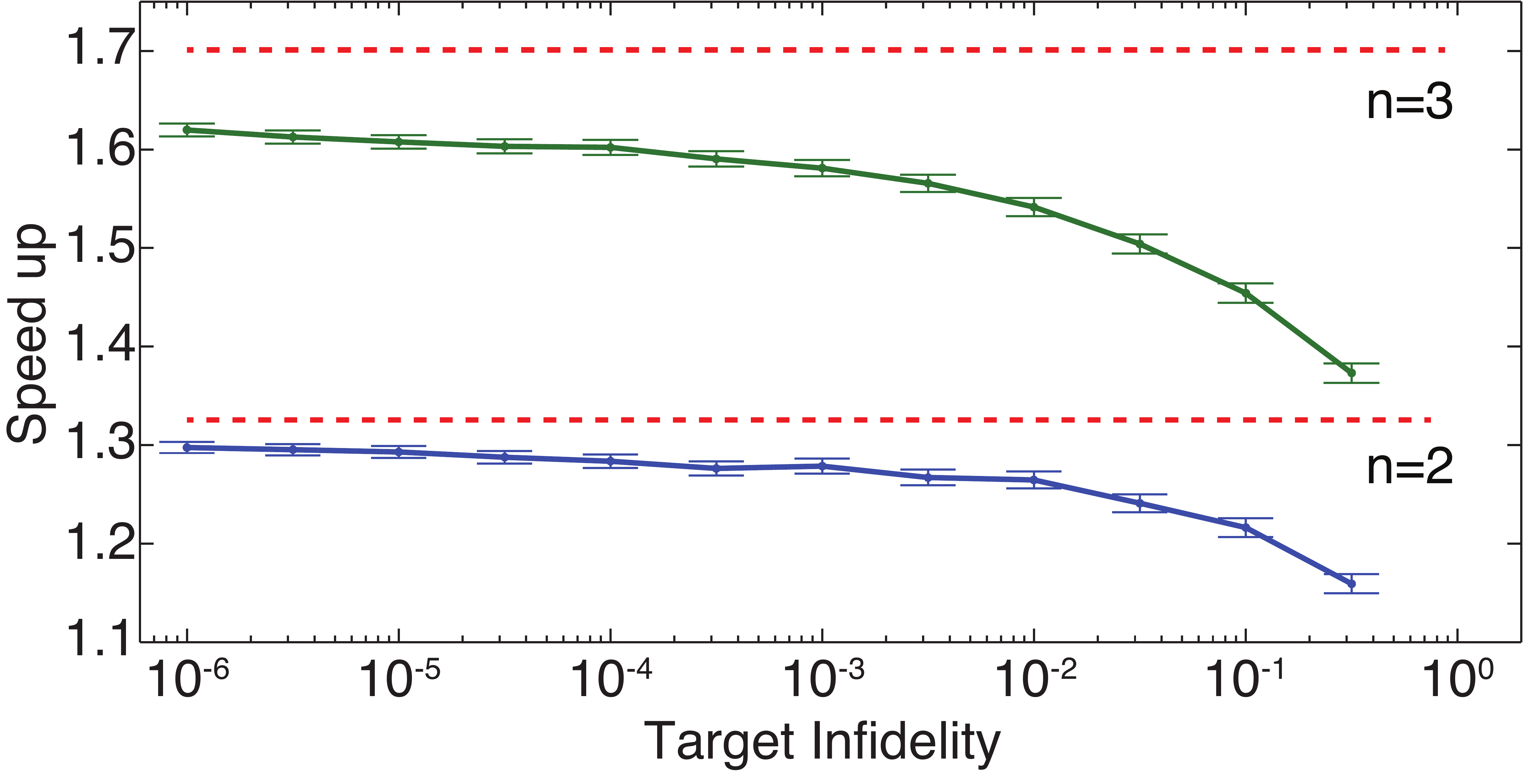}%.eps
\caption{The speed-up for a random permutation strategy for a register of $n$ qubits with (from top to bottom) $n=3$ (green) and $n=2$ (blue) where $\delta t = 6.25\times 10^{-4}\gamma^{-1}$. The dashed lines are the asymptotically calculated speedups. Averages were taken over 10,000 trajectories.}\label{Fig6} 
\end{center}
\end{figure} 

In order to determine the scaling of the speed up we must look at how the asymptotic speed up scales with $n$. Due to the difficulty of simulating large quantum systems we are restricted to $n\le5$. Time-asymptotic numerical results (see \frf{Fig7}) indicate that random permutations in the logical basis give a similar improvement to the Hamming-ordered feedback scheme in. The fit shown is
\begin{equation}
S = 0.397n+0.53,
\end{equation}
as, unlike in the case of the Hamming-ordered feedback scheme, a strictly proportional fit is not adequate. %Fitting the slope alone gives $S = 0.525n$.
\begin{figure}[h]
\begin{center}
\leavevmode \includegraphics[width=\hsize]{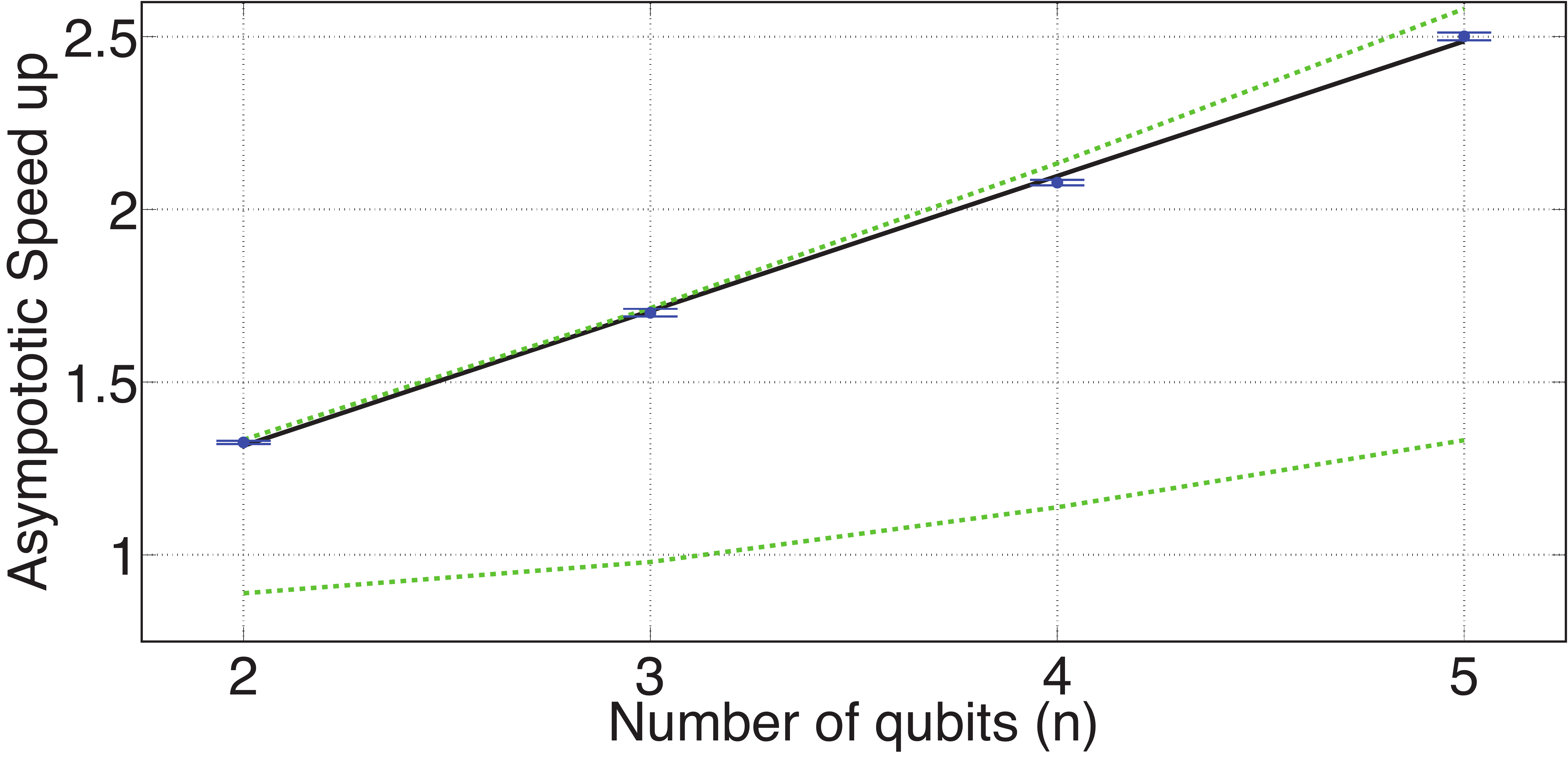}%.eps
\caption{The asymptotic speed-up in the mean time for a quantum register as a function of the number of qubits in the register. The numerically calculated asymptotic speed up are the data points with one standard deviation error bars. The dashed lines are the upper and lower bounds on the speed up given by \erf{RPLUBreg}. The solid line is the linear fit $S = 0.397n+0.53$.}\label{Fig7} 
\end{center}
\end{figure} 

The  fact that the numerically obtained speed-up is closer to the upper bound on \erf{RPLUBregsimp} than the lower bound can be understood by the following qualitative analysis. The speed-up in the measurement comes from making the large eigenvalues of $\rho$ more distinguishable \cite{ComWisJac08}.   This can be related to the Hamming distance between the two largest eigenvalues of $\rho$. Now consider the Hamming distance between two binary strings, call it $d_{\rm H}$, chosen at random with replacement (without replacement is asymptotically the same). The average distance is $\enavg{d_{\rm H}} \sim 0.5n$ \cite{FuKloShe99} which is precisely the upper-bound of \erf{RPLUBregsimp}. Clearly we are not quite achieving that bound for small $n$.

%==========================================================
\section{Discussion}\label{dis}
%==========================================================
We have shown that performing random permutations on a register of qubits increases the effective measurement strength and thus collapses the system into an eigenstate faster.

Because our open loop control procedures are effectively permutations of the eigenstates of the observed quantity, the control commutes with the measurement at all times. So while we have assumed our initial state was initially diagonal in the logical basis (the maximally mixed state) our protocol will work for any state. Specifically, at the end of the protocol one can calculationally retrodict the result of the unpermuted measurement. In some sense we can answer the counterfactual question ``had I not performed the control what would my measurement result be?" Consequently our protocol is compatable with state and process estimation protocols such as tomography.

It seems possible that there exists deterministic strategies which may achieve approximate asymptotic speed ups similar to those reported here. For $n=2$ we have found a single permutation, $P_{3124}$, which performs as well as the random strategy. The reason is simple, if the initial state is $\rho={\rm diag}(\lambda_0,\lambda_1,\lambda_2,\lambda_3)$ the permutation maps $\rho$ as follows ${\rm diag}(\lambda_0,\lambda_1,\lambda_2,\lambda_3)\mapsto {\rm diag}(\lambda_1,\lambda_2,\lambda_0,\lambda_3)\mapsto {\rm diag}(\lambda_2,\lambda_0,\lambda_1,\lambda_3)\mapsto{\rm diag}(\lambda_0,\lambda_1,\lambda_2,\lambda_3)$. When visualized on the Hamming cube this permutation makes all possible combinations of pairs of eigenvalues maximally distance at least once in the cycle. For larger registers it may be possible to find sequences with a small number of permutations that produce similar results. This is an open and interesting question for future research  and would drastically reduce the requirements of our scheme.

Finally, and perhaps most importantly, we suggest that future work should include imperfections in the control protocols as was done in Ref.~\cite{ComWis11b}, such as finite number of permutations in a fixed time interval, to see the advantage of our protocol persists.

\acknowledgments The authors acknowledge fruitful discussions with Kurt Jacobs and Carl Caves as well as discussions with Steve Bartlett and Peter Turner on averaging over the permutation group. This research was conducted by the Australian Research Council Centre of Excellence for Quantum Computation and Communication Technology (project number CE110001027).  JC also acknowledges support from the National Science Foundation Grant No. PHY-1212445 and from the Office of Naval Research Grant No. N00014-11-1-0082.

%================================================
\appendix
%================================================

\section{bounds on \erf{loginfide}}\label{appbound}

%----------------------------------------
\subsection{Upper bound}
%----------------------------------------
Recall that the operator $Z^r= I^{(1)}\otimes I^{(2)}\otimes \ldots \sigma_z^{(r)}\ldots \otimes I^{(n)}$ was transformed to $Z_r\mapsto Z_r-I$ and then we defined $Z^r_{s}=P_s Z^rP_s$ where $P_s$ is an element of permutation group ${\mathfrak P}(2^{n})$. Now consider the  $D\times D$ ($D=2^n$) matrix representation of $Z^r_{s}$
\begin{align}
Z^r_{s}=
\left(\begin{array}{cccc}
Z^r_{s,11} & 0 & 0 & \cdots\\
0 & Z^r_{s,22} & 0 & \cdots \\
0 & 0 & Z^r_{s,33}& \cdots \\
\vdots & \vdots & \vdots & \ddots\end{array}\right)
\end{align}
where $Z^r_{s,ij}\in\{0,-2\}$ is the element in the $i$'th row and $j$'th column of $Z^r_{s}$.
Using the state $\rho_{2}=\mathrm{diag}(1-\Delta,\Delta,0,\cdots,0)$ we see that
\begin{equation}
  \enavg{d\ln(1- \mathbb{P}_{0})}=-\frac{4\gamma \,dt}{D!}\frac{(1-\Delta)^2} {\Delta^2}\sum_{s}\sum _{r}\expt{Z^r_{s}}^2
\end{equation}
becomes
\begin{align}
  \enavg{d\ln(1- \mathbb{P}_{0})}=&-\frac{4\gamma \,dt}{D!}\frac{(1-\Delta)^2} {\Delta^2}\nonumber\\
  &\,\, \times\sum_{s}\sum _{r}[Z^r_{s,11}(1-\Delta) + Z^r_{s,22}\Delta]^2.
\end{align}
In the expansion of the sum we only keep terms of order $\Delta^2$ because of $(1/\Delta^2)$ out the front
\begin{align}
  \enavg{d\ln(1- \mathbb{P}_{0})}\approx&-\frac{4\gamma \,dt}{D!}\sum_{s}\sum _{r}\nonumber\\
  & [(Z^r_{s,11})^2 - 2Z^r_{s,11}Z^r_{s,22}  + (Z^r_{s,22})^2].
\end{align}

To make further progress we have two sums to work out. The first sum is
\begin{align}\label{usefull1}
\sum_{s}(Z^r_{s,ii})^2 = \frac{4}{2}D!=2D!\quad \forall r\,\, {\rm and}\,\, i.
\end{align}
%which is simply the statement that half of the time the element is $0$ and half of the time it is $4$ the average is $2$ which is then summed $D!$ times. 
The second sum is 
\begin{align}\label{usefull2}
\sum_{s}Z^r_{s,ii}Z^r_{s,jj} &= 4\frac{D}{2}\left ( \frac{D}{2}-1\right) (D-2)!\nn\\
&=\left [1- \frac{1}{D-1} \right ]D!\quad \forall r\,\, {\rm and}\,\, i\neq j.
\end{align}
%This can be understood as there are $D/2$ ways to choose the first element not to be zero i.e. $4$. Then for the second element  ... how does the logic work here. This gives $4(D/2)(D-1)(D-2)!$ as required.\\
Using those expressions 
\begin{align}
  \enavg{d\ln(1- \mathbb{P}_{0})}=&-8\gamma \,dt\sum _{r} \frac{D}{D-1}\\
  =&-8\gamma n\,dt \frac{D}{D-1}\\
=&-8\gamma n\,dt \frac{2^n}{2^n-1}\\
=&-16\gamma n\,dt \frac{2^{n-1}}{2^n-1}.
\end{align}

%----------------------------------------
\subsection{Lower bound}
%----------------------------------------
The lower bound on \erf{loginfide} is obtained by the same procedure as above except we use the state $\rho_{F}=(1-\Delta, \delta, \cdots, \delta)$ where $\delta = \Delta/(D-1)$, instead of $\rho_2$. Now the important part is
\begin{align}
&\sum_{s}\sum _{r}\expt{Z^r_{s}}^2=\nonumber\\
&\sum_{s}\sum _{r} \Big[  Z^r_{s,11}(1-\Delta) + Z^r_{s,22}\delta+\ldots + Z^r_{s,nn}\delta \Big]^2.
\end{align}
When expanding this sum we use \erf{usefull1}  and \erf{usefull2} as well as neglecting terms that do not have order $\Delta^2$ to obtain
\begin{align}
&\sum_{s}\sum _{r}\expt{Z^r_{s}}^2\approx \sum_{s}\sum _{r} \Big \{  [\Delta^2  +(D-1)\delta^2 ](Z^r_{s,11})^2+\nonumber\\
&\,\,[(D-1)(D-2)\delta^2 -2(D-1)\Delta \delta ]Z^r_{s,11}Z^r_{s,22} \Big\} .
\end{align}
Summing over permutations and then simplifying we find
\begin{align}
\sum _{r}\expt{Z^r_{s}}^2 &= \Delta^2 \sum _{r} \Big \{  \frac{D+2}{D-1}-\frac{D-2}{(D-1)^2} \Big\} \\
%&= n \Big \{  \frac{D+2}{D-1}-\frac{D-2}{(D-1)^2} \Big\} \\
&= \Delta^2 n \Big \{  \frac{2^n+2}{2^n-1}-\frac{2^n-2}{(2^n-1)^2} \Big\} \\
&=  \Delta^2 n \frac{2^{2n}}{\left(2^n-1\right)^2}
\end{align}
Finally we obtain
\begin{align}
\enavg{d\ln(1- \mathbb{P}_{0})}=&-4\gamma n \,dt \frac{2^{2n}}{\left(2^n-1\right)^2}\\
=& -16\gamma  \,dt \frac{n}{4}\frac{2^{2n}}{\left(2^n-1\right)^2}.
\end{align}

\end{document}